\documentclass[11pt]{article}
\usepackage{amsmath}
\newcommand{\p}{{\cal P}}
\newcommand{\ctp}[2]{\tilde{{\cal P}}^{#1}_{#2}}
\newcommand{\g}[2]{\Gamma^{#1}_{#2}}
\newcommand{\h}[1]{{\cal H}^{#1}}
\newcommand{\sw}[2]{\begin{array}{c} #1 \\ #2 \end{array}}
\newcommand{\ket}[1]{\mid #1\rangle}
\newcommand{\bra}[1]{\langle #1 \mid}
\newtheorem{The}{Theorem}[section]

\begin{document}
\title{SEARCHING FOR NEW CONDITIONS FOR FERMION $N$-REPRESENTABILITY}
\author{Hubert Grudzi\'nski \\  Department of Physics, Academy of Bydgoszcz, \\
85-072 Bydgoszcz, pl. Weyssenhoffa 11, Poland \\( e-mail: hubertg@ab-byd.edu.pl)\\
Jacek Hirsch \\ Institute of Physics, Nicholas Copernicus University,\\
87-100 Toru\'n, Poland\\     
(e-mail: jacekh@phys.uni.torun.pl)} 
\date{}
\maketitle
\begin{abstract}
New elements of the dual cone of the set of fermion $N$-representable
2-density operators are proposed. So far, the explicit form of the 
corresponding necessary conditions for $N$-representability is obtained for $N=3$.
In this case the new condition is stronger than the known B- and C-conditions
for 3-representability. The results provide evidence that in the spectral 
decomposition of the $N$-representable 2-density operator there exists an intrinsic 
relation between the eigenvalue and the corresponding eigenfunction.
\end{abstract}
{\bf keywords: fermion $N$-representability problem, conditions for 
$N$-representability}
\section{Introduction}

The $N$-representability problem \cite{AJC1,AJC5,GP,Kuhn,HK1,Low,McW} appears in non-relativistic quantum
mechanics of $N$-fermion systems. The Hamiltonian $H^N$ for systems of
$N$-fermions contain only operators with 1- and 2-body interactions:
$$
H^N = \sum_{i=1}^{N}H^1(i) + \sum_{1 \leq i < j \leq N} H^2(i,j) =
\sum_{1 \leq i < j \leq N} h^2(i,j) .
$$
The ground state energy of the system can be determined variationally by 
minimizing the $N$-particle functional
$$
E = 
\begin{array}[t]{c}
inf \\ D^N \in \p^N 
\end{array}
Tr(H^N D^N)
$$
over the set $\p^N$ consisting of all fermion $N$-particle density 
operators. Instead, because of the appearance of at most 2-body interactions
between the particles, the ground state energy of the system of $N$-fermions
could be, in principle, determined variationally by minimizing the 2-particle
functional
$$
E = 
\begin{array}[t]{c}
inf \\ D^2 \in \p^2_N
\end{array}
\binom{N}{2} Tr(h^2 D^2)
$$
over the set $\p^2_N$ consisting of all fermion 2-particle reduced density
operators, i.e. such 2-particle density operators $D^2$ which possess an
$N$-particle fermion preimage $D^N$:
$$
D^2 = L^2_ND^N = Tr_{3,\ldots,N} D^N .
$$
It is known \cite{AJC1,HK1}  that the set $\p^2_N$ is a proper convex subset of $\p^2$,
the set of all fermion 2-density operators. However, the complete characterization
of $\p^2_N$ as a convex proper subset of $\p^2$ is not yet known. It has been
shown \cite{HK1} that a knowledge of all exposed points (which are extreme) of $\p^2_N$
is sufficient to characterize the closure of $\p^2_N$; only some of them are known. 
The dual characterization of $\p^2_N$ involves a determination of the
dual (polar) cone $\ctp{2}{N}$ consisting of 2-particle self-adjoint operators 
$b^2$ for which $Tr(b^2 D^2)\geq 0$, $ \forall D^2 \in \p^2_N $, that is 
equivalent to the positive-semidefiniteness of the $N$-particle operator
$\g{N}{2} b^2 =b^2\wedge I^{\wedge (N-2)} = A^N b^2 \otimes I^{\otimes (N-2)}
A^N \geq 0$. The dual cone $\ctp{2}{N}$ is a convex one. Any element of the dual
cone $\ctp{2}{N}$ provides an $N$-representability condition. Those coming from
the extreme elements of $\ctp{2}{N}$ are the strongest ones. They give the
hyperplane characterization of $\p^2_N$ and thus the solution of the
$N$-representability problem. Several necessary conditions for $N$-representability
have been derived and some of their structural features and mutual interrelations
are established \cite{AJC2,AJC3,ERD1,RME1,RME2,RME3,RME4,RME5,GP,HG1,HG2,HG3,HG4,
HK1,HK2,HK3,HK4,McR,WW,YK}.

In this paper we find  new elements of the dual cone $\ctp{2}{N}$, and thus
define new necessary conditions for $N$-representability of a trial 2-particle
density matrix $D^2$. In the case N=3, in which the new elements of the dual cone
are given explicitly, they lead to necessary conditions for 3-representability
of the trial $D^2$ which are stronger in comparison with the known $B$ and $C$
conditions \cite{AJC3,AJC4,HK3}. The results show that in the spectral 
decomposition of a 2-particle fermion density operator $D^2$ there exists an 
intimate relation between the eigenvalue and the corresponding eigenfunction 
 which has to be satisfied in order that $D^2$ can be 
3-representable ( in general $N$-representable). More precisely, the condition
obtained for 3-representability says that the upper bound of the eigenvalue 
of $D^2$ is a functional of the corresponding eigenfunction. It is worthwhile
to remember that for $N$-representability of a 1-particle density matrix $D^1$
the upper bound on the eigenvalues does not depend on the eigenfunctions.
This explains why the $N$-representability problem for $D^2$ is so much harder
than for $D^1$.
\section{The dual P-condition.}

In this paper the underlying 1-particle Hilbert space $\h{1}$  is finite 
dimensional $dim \h{1} =n$. $ \h{\wedge 2}$ is a 2-particle antisymmetric Hilbert
space, the Grassmann product $\h{1}\wedge \h{1} = A^2\h{1}\otimes\h{1}$.
$P^2_g = g^2\otimes \bar{g}^2$ is the projection operator onto a 2-particle
antisymmetric function $g^2\in\h{\wedge 2}$.													
The operator 
$\binom{N}{2} P^2_g\wedge I^{\wedge (N-2)} = \binom{N}{2} A^N P^2_g \otimes I^{
\otimes (N-2)} A^N = A^N \sum_{1\leq i<j \leq N} P^2_g(i,j) \otimes I^{N-2}(1,\ldots
,i-1,i+1,\ldots ,j-1,j+1,\ldots ,N) A^N$
is the simplest (elementary) antisymmetric operator with "2-body interactions"
acting on $\h{\wedge N}$. Because $\binom{N}{2} P^2_g \wedge I^{\wedge (N-2)}$
is positive-semidefinite, $P^2_g$ belongs to the dual cone $\ctp{2}{N}$, and
therefore $Tr(D^2P^2_g) \geq 0$ is a necessary condition for $N$-representability
of $D^2$ (the P-condition). Besides, there exists another element of 
$\ctp{2}{N}$ that is generated by the 2-particle
antisymmetric function $g^2\in\h{\wedge 2}$. Let $\Lambda_{max}(g^2)$ denote
the maximal eigenvalue of $\binom{N}{2} P^2_g \wedge I^{\wedge(N-2)}$, which in
general depends on $g^2$, then the operator $\Lambda_{max}(g^2) I^{\wedge N} -
\binom{N}{2} P^2_g \wedge I^{\wedge(N-2)} \geq 0 $ (is positive-semidefinite),
and therefore the 2-particle operator $\frac{\Lambda_{max}(g^2)}{\binom{N}{2}}
I^{\wedge 2} - P^2_g$ belongs to the dual cone $ \ctp{2}{N}$ . Thus, we have 
obtained
\begin{The}
For any $g^2\in\h{\wedge 2}$ the operator
\begin{equation}
\label{dPc0}
\frac{\Lambda_{max}(g^2)}{\binom{N}{2}} I^{\wedge 2} - P^2_g  \in \ctp{2}{N}, 
\end{equation}
and gives the following necessary condition for 
$N$- representability of a 2-particle fermion density matrix $D^2$:
$Tr[(\frac{\Lambda_{max}(g^2)}{\binom{N}{2}} I^{\wedge 2} - P^2_g) D^2]\geq 0$, i.e.
\begin{eqnarray}
\label{dPc1}
Tr(D^2 P^2_g) \leq \frac{\Lambda_{max}(g^2)}{\binom{N}{2}}, & \forall g^2
\in\h{\wedge 2}.
\end{eqnarray}
Here, $\Lambda_{max}(g^2)$ is the maximal eigenvalue of the
operator $\binom{N}{2} P^2_g\wedge I^{\wedge(N-2)}$.
\end{The}
We propose to call the new condition "the dual P-condition".

In particular, from the above theorem  follows
\begin{The}
If $D^2 = \sum_{i=1}^{\binom{n}{2}} \lambda_i P^2_{g_i} $ is the spectral
decomposition of $D^2$, then it is a necessary condition for $N$-representability
that the eigenvalues $\lambda_i $ must satisfy the inequalities
$\lambda_i \leq \frac{\Lambda_{max}(g^2_i)}{\binom{N}{2}}$. Here, $\Lambda_{max}
(g^2_i)$ is the maximal eigenvalue of the operator $\binom{N}{2} P^2_{g_i}\wedge
I^{\wedge (N-2)}$, where $g^2_i$ is the eigenfunction corresponding to the
eigenvalue $\lambda_i$.
\end{The}

This theorem shows that the bound on the eigenvalue is a functional of the
corresponding eigenfunction. For the $N$-fermion 1-body elementary operator
$N P^1_g \wedge I^{\wedge(N-1)}$, $g^1 \in \h{1}$, the maximal eigenvalue
$\Lambda_{max}(g^1) = 1$ for any $g^1$, because $N P^1_g \wedge I^{\wedge(N-1)}$
is a projection operator. Hence, the equivalent of Theorem 2.1 says that
$Tr(D^1 P^1_g) \leq \frac{1}{N}$ for arbitrary $g^1 \in \h{\wedge1}$, which
in particular means that the bound on an eigenvalue of an $N$-representable
1-density operator does not depend on the corresponding eigenfunction, 
contrary to the 2-density matrix case. 

As  seen from Theorem 2.1, the new necessary condition for $N$-representability
of a 2-particle density matrix $D^2$ requires  knowledge of the maximal eigenvalue 
$\Lambda_{max}(g^2)$ of the $N$- particle operator $\binom{N}{2} P^2_g \wedge I^{
\wedge(N-2)}$. It is rather hopeless to find $\Lambda_{max}(g^2)$ for arbitrary
$N$ and any $g^2 \in \h{\wedge 2}$. But even the solution for some "simple"
$g^2$ would contribute to  knowledge of the structure of both the dual cone
$\ctp{2}{N}$ and the set $\p^2_N$ of fermion $N$-representable 2-density 
operators. So far, we have results for $N=3$, in which case it was possible to
find the spectral decomposition of the elementary operator $3 P^2_g \wedge I^1$
for arbitrary $g^2 \in \h{\wedge 2}$. The details and proofs concerning this
spectral decomposition and the reduced 2- and 1-particle density operators
corresponding to the eigenstates of the operator $3 P^2_g \wedge I^1$ 
will be published in a separate paper .
In this paper we use only the following result which we formulate as

\begin{The}
Let $\h{1}$ be a finite dimensional Hilbert space (dim $\h{1} = n$), and
$\h{\wedge 2} = \h{1} \wedge \h{1}$ denotes the 2-particle antisymmetric space
generated by $\h{1}$ (the 2-fold Grassmann product of $\h{1}$). Let $P^2_g$
denote the 1-dimensional projection operator onto a 2-particle antisymmetric
function $g^2 \in \h{\wedge 2}$ of 1-rank $r=2s$ possessing the 
canonical decomposition $g^2 = \sum_{i=1}^{s} \xi _i \ket{2i-1,2i}$ with
$\sum_{i=1}^{s} \mid \xi_i\mid^ 2 = 1$, where $\ket{2i-1,2i} = \sqrt{2} \phi ^1
_{2i-1} \wedge \phi ^1 _{2i} = \frac{1}{\sqrt{2}} det(\phi^1_{2i-1},\phi^1_{2i})$
is the 2-particle normalized Slater detrminant. Let the identity operator $I^1$
on $\h{1}$ possess the decomposition $I^1 = \sum_{i=1}^{r=2s} P^1_i + \sum_{i=r+1}^{n} P^1_i$, 
where $P^1_i =\ket{i}\bra{i} = \phi ^1_i \otimes \bar{\phi}^1_i (i=1,\ldots ,n)$
are 1-dimensional mutually orthogonal projection operators onto the functions 
$\ket{i} = \phi^1_i$. Then, the 3-particle operator $3 P^2_g \wedge I^1$ 
possesses the following spectral decomposition 
$$
3 P^2_g \wedge I^1 = \sum_{k=1}^{s=r/2} (1 - \mid \xi_k \mid ^2) (P^3_{g_{2k-1}} +
P^3_{g_{2k}}) + \sum_{l=r+1}^{n} P^3_{g_l}  + 0 \cdot Ker( 3 P^2_g \wedge I^1).
$$ 
Here, $P^3_{g_{2k-1}}, P^3_{g_{2k}} (k=1,\ldots ,s=r/2), P^3_{g_l} (l=r+1,\ldots ,
n)$ are 1-dim projectors onto the following functions:
$$
g^3_{2k-1} = \frac{1}{\sqrt{1-\mid\xi_k \mid ^2}} \sum_{\sw{i=1}{(i\ne k)}}^{s} 
\xi_i\ket{2i-1,2i,2k-1} = \sqrt{\frac{3}{1-\mid\xi_k\mid^2}} g^2 \wedge \ket{2k-1},
$$ 
$$
g^3_{2k} = \frac{1}{\sqrt{1-\mid\xi_k \mid ^2}}\sum_{\sw{i=1}{(i\ne k)}}^{s}
\xi_i\ket{2i-1,2i,2k} = \sqrt{\frac{3}{1-\mid\xi_k\mid^2}} g^2 \wedge\ket{2k},
$$
$$
g^3_l = \sum_{i=1}^{s} \xi_i\ket{2i-1,2i,l} = \sqrt{3} g^2\wedge \ket{l},
$$
$Ker(3 P^2_g \wedge I^1)$ denotes the projection operator onto the null-space
of the operator $3 P^2_g \wedge I^1$, which is of dimension $\binom{n}{3}-n$.
The symbols of the type $\ket{2i-1,2i,2k-1}$ denote the appropriate 3-particle
Slater determinants.
\end{The}

Now, we can apply results of Theorem 2.3 to find new necessary conditions for
3-representability according to the formulae of Theorems 2.1 and 2.2 .
The required $\Lambda_{max}(g^2)$ is either $1-\mid\xi\mid ^2_{min}$, if the
1-rank of $g^2$ is equal to the dimension $n$ of the 1-particle Hilbert space
$\h{1}$ ($r=2s=n$), or $\Lambda_{max}(g^2) = 1$, if $r$ is less than $n$
($r<n$). Hence, we have 
\begin{The}
If $D^2$ ($Tr D^2 = 1$) is a 2-fermion density matrix, then for 3-representability
it must satisfy  for any $g^2 \in\h{\wedge 2}$, with the canonical
decomposition $g^2 = \sum_{i=1}^{r/2}\xi_i \ket{2i-1,2i}$, the following inequality:
\begin{equation}
\label{dPc2}
Tr(D^2P^2_g) \leq \frac{1}{3} (1-\mid\xi\mid^2_{min}),
\end{equation}
if the 1-rank of $g^2$ equals $n=dim\h{1}  (r=n), $ 
\begin{equation}
\label{dPc3}
Tr(D^2P^2_g) \leq \frac{1}{3},
\end{equation}
if the 1-rank $r$ of $g^2$ is less than $ dim \h{1}  (r<n).$
\end{The}

In particular, if we take as $g^2$ the eigenfunctions $g^2_i$ of $D^2 $, we get
\begin{The}
If $D^2$ has the spectral decomposition $D^2 = \sum_{i=1}^{\binom{n}{2}}\lambda_i
P^2_{g_i}$ with $g^2_i = \sum_{j=1}^{s_i}\xi_{ij} \ket{2j-1,2j}$,
then $\lambda_i \leq \frac{1}{3}(1-\mid\xi_{ij}
\mid^2_{\tiny\begin{array}{c} min\\j\end{array}})$, if the
eigenfunction $g^2_i$ has 1-rank $r_i =2s_i=n$, and $\lambda_i \leq\frac{1}{3}$,
if the eigenfunction belonging to $\lambda_i$ has 1-rank $r_i<n=dim\h{1}$.
\end{The}

It is worthwhile to observe that the eigenfunctions belonging to the maximal
eigenvalue $\lambda = \frac{1}{3}$ cannot have  full 1-rank ( $ r=n=dim 
\h{1}$) if  $D^2$ is 3-representable, and on the other hand, the eigenvalues
corresponding to the eigenfunctions with full 1-rank  must be strictly less than
$\frac{1}{3}$.
\section{Strength of the dual P-condition for fermion 3-representability}
\setcounter{equation}{0}

It is important to compare the new condition with the already known conditions 
for 3-representability because the effort to find $\Lambda_{max}(g^2)$ for at 
least some $g^2$ in the general case (arbitrary $N$) might not pay off. Since the
new condition is an estimation from the above on the expectation value of $D^2$ in
any state $g^2 \in \h{\wedge 2}$, it can be compared with the $B$- and $C$-
conditions for $N$-representability of $D^2$ \cite{AJC2,AJC3,HK3}:
\begin{eqnarray}
\label{BaC1}
Tr[B^2_N(g^2) D^2] \geq 0, & Tr[C^2_N(g^2)D^2] \geq 0, & 
\forall g^2 \in \h{\wedge 2},
\end{eqnarray} 
where
$$
B^2_N(g^2) = I^{\wedge 2}-(N-2) L^1_2 P^2_g\wedge I^1-(N-1)P^2_g \in \ctp{2}{N},
$$
$$
C^2_N(g^2) = (n-N+2) L^1_2P^2_g\wedge I^1- (N-1)P^2_g \in \ctp{2}{N}.
$$
In the case $N=3$,  conditions (3.1) take the form:
\begin{eqnarray}
\label{BaC2}
Tr(D^2P^2_g)\leq \frac{1}{2} [1-Tr(L^1_2D^2 L^1_2P^2_g)], & \forall g^2 \in \h{
\wedge 2}
\end{eqnarray} 
\begin{eqnarray}
\label{BaC3}
Tr(D^2P^2_g)\leq 
\frac{n-1}{2} Tr(L^1_2D^2 L^1_2P^2_g), & \forall g^2 \in \h{\wedge 2}.
\end{eqnarray}

In order to compare the inequalities (2.2), (3.2), (3.3), we choose as $g^2$ the
so called "extreme geminal" \cite{AJC2} $g^2_{extr}$ of 1-rank $r=2s=n=dim\h{1}$ possessing
the following canonical decomposition: 
$$
g^2_{extr} = \sum_{i=1}^{n/2}\sqrt{\frac{2}{n}}\ket{2i-1,2i}. 
$$ 
The required by inequality (\ref{dPc2}) $\mid \xi \mid^2_{min} = \frac{2}{n}$, and 
therefore the new condition gives 
\begin{equation}
\label{dPc4}
Tr(D^2P^2_{g_{extr}})\leq \frac{1}{3}(1-\frac{2}{n}) .
\end{equation}
On the other hand, since $L^1_2P^2_{g_{extr}}= \sum_{i=1}^{n} \frac{1}{n} P^1_i$
, $P^1_i = \ket{i} \bra{i}$, $\sum_{i=1}^{n} P^1_i = I^1 $, we have from both
(\ref{BaC2}) and (\ref{BaC3}) the same result
\begin{equation}
\label{BaC4}
Tr(D^2P^2_{g_{extr}}) \leq \frac{1}{2}(1-\frac{1}{n}).
\end{equation}
Comparing (\ref{dPc4}) with (\ref{BaC4}), we see that the new condition is stronger than the 
B and C ones. So far, we were unable to establish the relation between the new
condition (\ref{dPc2}) and the G-condition \cite{GP} for 3-representability. This would be
important because it is known \cite{RME5,HK2} that the G-condition implies the B- and C- ones.
\section{The strengthened B-condition}
\setcounter{equation}{0}
While comparing the new condition with the B-condition we found out that the
B-condition could  in principle be improved for N odd. $B^2_N(g^2) \in \ctp{2}{N}$,
$\forall g^2 \in \h{\wedge 2}$ means that the operator $\binom{N}{2}  B^2_N(g^2)
\wedge I^{\wedge(N-2)} \geq 0$, $\forall g^2 \in \h{\wedge 2}$. 
If $\Lambda_{min}^{B(g)}$ is the minimal eigenvalue of the operator
$\binom{N}{2} B^2_N(g^2) \wedge I^{\wedge(N-2)}$, then the operator
$\binom{N}{2} B^2_N (g^2) \wedge I^{\wedge(N-2)} - \Lambda_{min}^{B(g)} I^{\wedge N}$
 is positive-semidefinite and therefore 
 \begin{equation}
 \label{sBc0}
 \binom{N}{2}B^2_N(g^2) - \Lambda_{min}^{B(g)} I^{\wedge 2}  \in\ctp{2}{N}, 
 \end{equation}
 giving a necessary condition for $N$-representability (the strengthened 
 B-condition). It is known \cite{AJC3} that for $N$ even $\Lambda_{min}^{B(g)} =0$, 
 $\forall g^2 \in \h{\wedge 2}$. For $N=3$ and $g^2_{extr}$, the eigenvalue 
 $\Lambda_{min}^{B(g_{extr})}$ is greater than zero, and can be found explicitly. 
 The operator 
 $$
 3B^2_3(g^2_{extr}) \wedge I^{1} = 3(1-\frac{1}{n}) I^{\wedge 3} - 
 2(3P^2_{g_{extr}} \wedge I^{1})
 $$
possesses the following minimal eigenvalue
$$
\Lambda_{min}^{B(g_{extr})} = 3(1-\frac{1}{n} ) - 2(1-\frac{2}{n})=1+\frac{1}{n},
$$
where Theorem 2.3 has been used. Hence, the strengthened B-condition in this case is
\begin{equation}
\label{sBc1}
3B^2_3(g_{extr}^2) - \Lambda_{min}^{B(g_{extr})} I^{\wedge 2}=\frac{2(n-2)}{n} 
I^{\wedge 2} - 2(3P^2_{g_{extr}}) \in\ctp{2}{N},
\end{equation}
and therefore 
\begin{equation}
\label{sBc2}
Tr(D^2P^2_{g_{extr}}) \leq \frac{1}{3} (1-\frac{2}{n}).
\end{equation}
Comparing inequalities (\ref{dPc4}) and (\ref{sBc2}) we see that both the new 
conditions for 3-representability in the case under consideration 
($g^2=g^2_{extr}$) give the same bound from above on the expectation value 
$Tr(D^2P^2_{g_{extr}})$.

It seems that the above results suggest, it would be worthwhile to make an effort 
to extend the new conditions to arbitrary $N$ for at least some $g^2$ for which 
the maximum eigenvalue $\Lambda_{max}(g^2)$ of the operator 
$\binom{N}{2}P^2_g \wedge I^{\wedge (N-2)}$ can be found, e.g. the extreme geminal 
$g^2_{extr}$. So far, we have obtained only partial information about the spectral
decomposition of the operator
$\binom{N}{2}P^2_{g_{extr}}\wedge I^{\wedge(N-2)}$ for arbitrary $N$. However, at 
least in this case it is realistic to succeed. The other known conditions for
$N$-representability could be treated in a similar way as the P- and B- conditions
considered in this paper, provided  the maximal and minimal eigenvalues of the
appropriate operators were known. The obtained results concerning arbitrary $N$ 
will be published later.
\vspace{0.5cm}

\noindent
{\bf Acknowledgements}
\vspace{0.5cm}

This work was supported in part by the Polish Committee of Scientific Research 
(KBN) under grant No. 5 P03B 083 20.
\end{document}